\documentclass[fleqn,12pt,twoside]{article}
\usepackage{espcrc1}

\usepackage{graphicx}
\usepackage[figuresright]{rotating}

\newcommand{\AmS}{{\protect\the\textfont2
  A\kern-.1667em\lower.5ex\hbox{M}\kern-.125emS}}

\title{Role of Large Gluonic Excitation Energy for Narrow Width of Penta-Quark Baryons in QCD String Theory}

\author{H.~Suganuma\address[TITech]{Tokyo Institute of Technology, Ohokayama 2-12-1, Meguro, Tokyo 152-8551, Japan
\vspace{-0.1cm}
},
F.~Okiharu\address[Nihon]{Dept. of Phys., Nihon University, Kanda-Surugadai 1-8-14, Chiyoda 101-8308, Japan
\vspace{-0.1cm}
},
T.T.~Takahashi\address[YITP]{Yukawa Institute for Theoretical Physics, Kyoto University, 
Kyoto 606-8502, Japan} 
and H.~Ichie\addressmark[TITech]}

\begin{document}

\maketitle

\begin{abstract}
We study the narrow decay width of low-lying penta-quark baryons in the QCD string theory
in terms of gluonic excitations. 
In the QCD string theory, the penta-quark baryon decays via a gluonic-excited state  
of a baryon and meson system, where a pair of Y-shaped junction and anti-junction is created.
Since lattice QCD shows that the lowest gluonic-excitation energy takes a 
large value of about 1 GeV, the decay of the penta-quark 
baryon near the threshold is considered as a quantum tunneling process via a highly-excited state 
(a gluonic-excited state) in the QCD string theory. This mechanism strongly suppresses the decay and leads to 
an extremely narrow decay width of the penta-quark system.
\end{abstract}

\section{3Q, 4Q, 5Q Potentials and Color-Flux-Tube Picture from Lattice QCD}

In 1969, Nambu first presented the string picture for hadrons \cite{N697074}. 
Since then, the string theory has provided many interesting ideas in 
the wide region of the particle physics.

Recently, various candidates of multi-quark hadrons (penta-quarks and tetra-quarks) 
have been experimentally observed \cite{Theta}.
As a remarkable feature of multi-quark hadrons, their decay widths are extremely narrow \cite{Z04}, 
which gives an interesting puzzle in the hadron physics.
In this paper, we study the narrow decay width of 
penta-quark baryons in the QCD string theory \cite{BKST04,STOI04}, with referring 
recent lattice QCD results \cite{STOI04,TS01,TS02,OST04,OST04p,TS03,TS04,IBSS03}.

First, we show the recent lattice QCD studies of the inter-quark potentials in 3Q, 4Q and 5Q 
systems \cite{STOI04,TS01,TS02,OST04,OST04p}, and 
revisit the color-flux-tube picture for hadrons.
For more than 300 different patterns of spatially-fixed 3Q systems, 
we perform accurate and detailed calculations for the 3Q potential 
in SU(3) lattice QCD with  
($\beta$=5.7, $12^3\times 24$),
($\beta$=5.8, $16^3\times 32$), 
($\beta$=6.0, $16^3\times 32$) and 
($\beta=6.2$, $24^4$), 
and find that the ground-state 3Q potential $V_{\rm 3Q}^{\rm g.s.}$
is well described by the Coulomb plus Y-type linear potential, i.e., Y-Ansatz,  
\begin{eqnarray}
V_{\rm 3Q}^{\rm g.s.}=-A_{\rm 3Q}\sum_{i<j}\frac1{|{\bf r}_i-{\bf r}_j|}+
\sigma_{\rm 3Q}L_{\rm min}+C_{\rm 3Q},
\end{eqnarray}
within 1\%-level deviation \cite{STOI04,TS01,TS02}.
Here,  $L_{\rm min}$ is the minimal value of the total length of the flux-tube, 
which is Y-shaped for the 3Q system.
To demonstrate this, we show in Fig.1(a) the 3Q confinement potential $V_{\rm 3Q}^{\rm conf}$, 
i.e., the 3Q potential subtracted by the Coulomb part, 
plotted against the Y-shaped flux-tube length $L_{\rm min}$.
For each $\beta$, clear linear correspondence is found between the 3Q confinement potential 
$V_{\rm 3Q}^{\rm conf}$ and $L_{\rm min}$, 
which indicates Y-Ansatz for the 3Q potential. 

Furthermore, a clear Y-type flux-tube formation is actually observed 
for the spatially-fixed 3Q system in lattice QCD \cite{STOI04,IBSS03}. 
Thus, together with recent several other analytical and numerical studies \cite{KS03,C04,BS04},
Y-Ansatz for the static 3Q potential seems to be almost settled. 
This result indicates the color-flux-tube picture for baryons.

\begin{figure}[h]
\vspace{-0.5cm}
\begin{center}
\includegraphics[height=3.8cm]{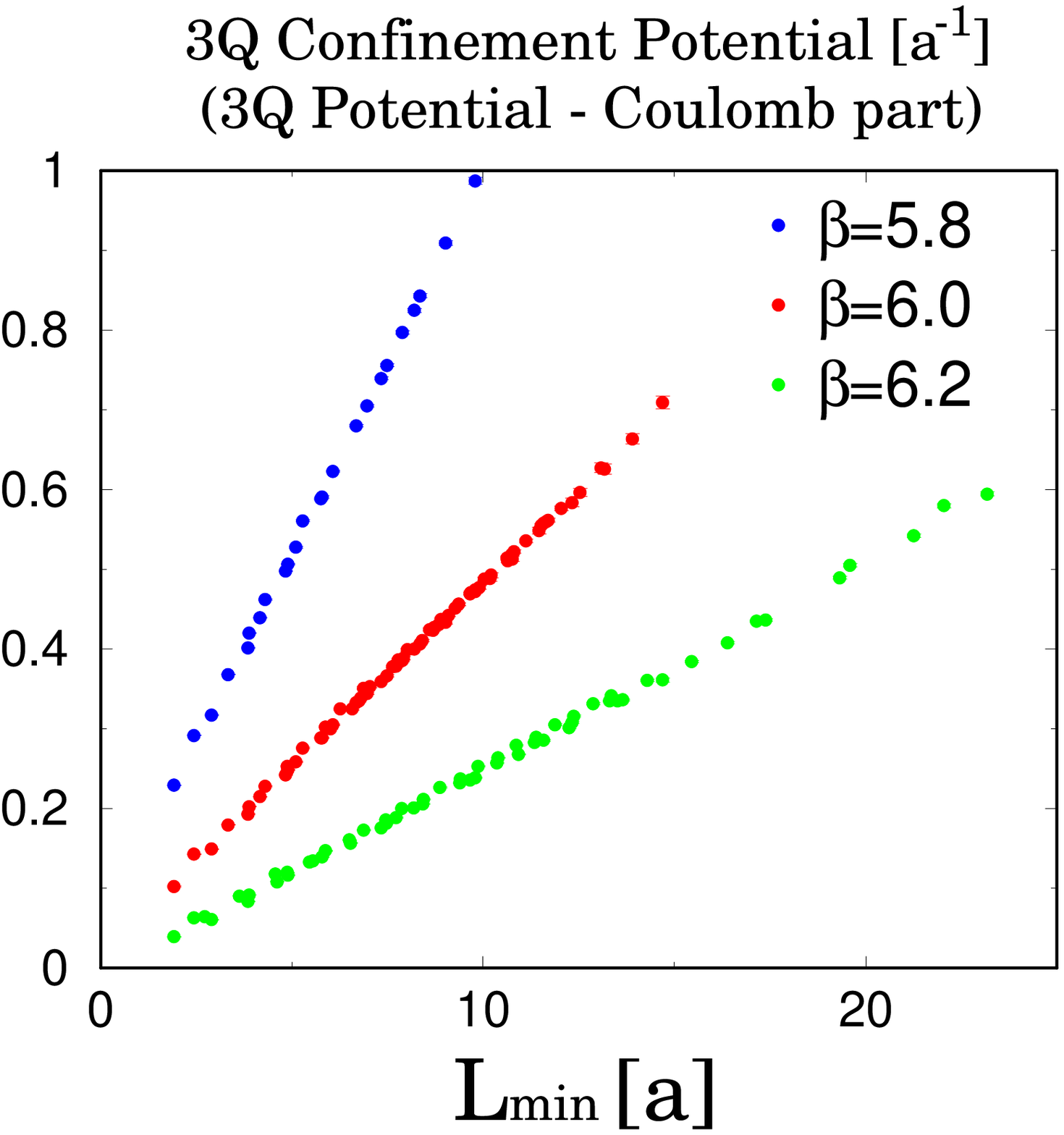}
\includegraphics[height=3.8cm]{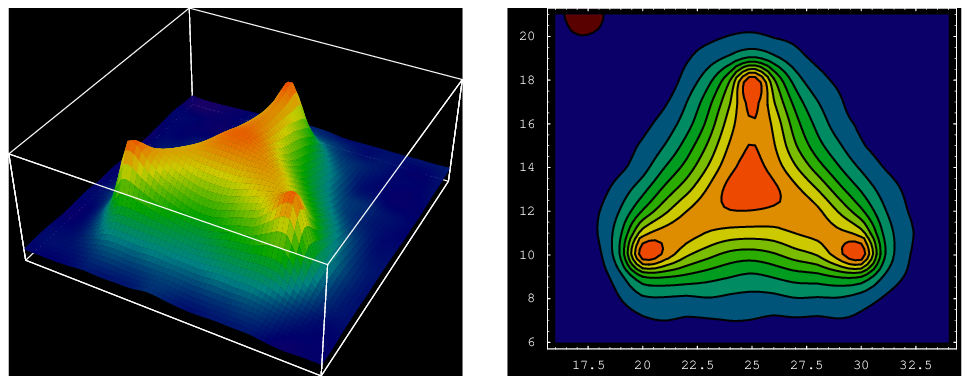}
\vspace{-0.7cm}
\caption{
(a) The 3Q confinement potential $V_{\rm 3Q}^{\rm conf}$, 
i.e., the 3Q potential subtracted by the Coulomb part, 
plotted against 
the Y-shaped flux-tube length $L_{\rm min}$ 
in the lattice unit.
(b) The lattice QCD result for Y-type flux-tube formation 
in the spatially-fixed 3Q system.
}
\end{center}
\vspace{-1cm}
\end{figure}

We perform also the first study of the multi-quark potentials in SU(3) lattice QCD \cite{STOI04,OST04,OST04p}, 
and find that they can be expressed as 
the sum of OGE Coulomb potentials and the linear potential based on the flux-tube picture. 
(This lattice result presents the proper Hamiltonian for the quark-model calculation of the multi-quark systems.)
In fact, the lattice QCD study indicates the color-flux-tube picture even for the multi-quark systems.

\begin{figure}[h]
\vspace{-0.6cm}
\begin{center}
\includegraphics[height=2.8cm]{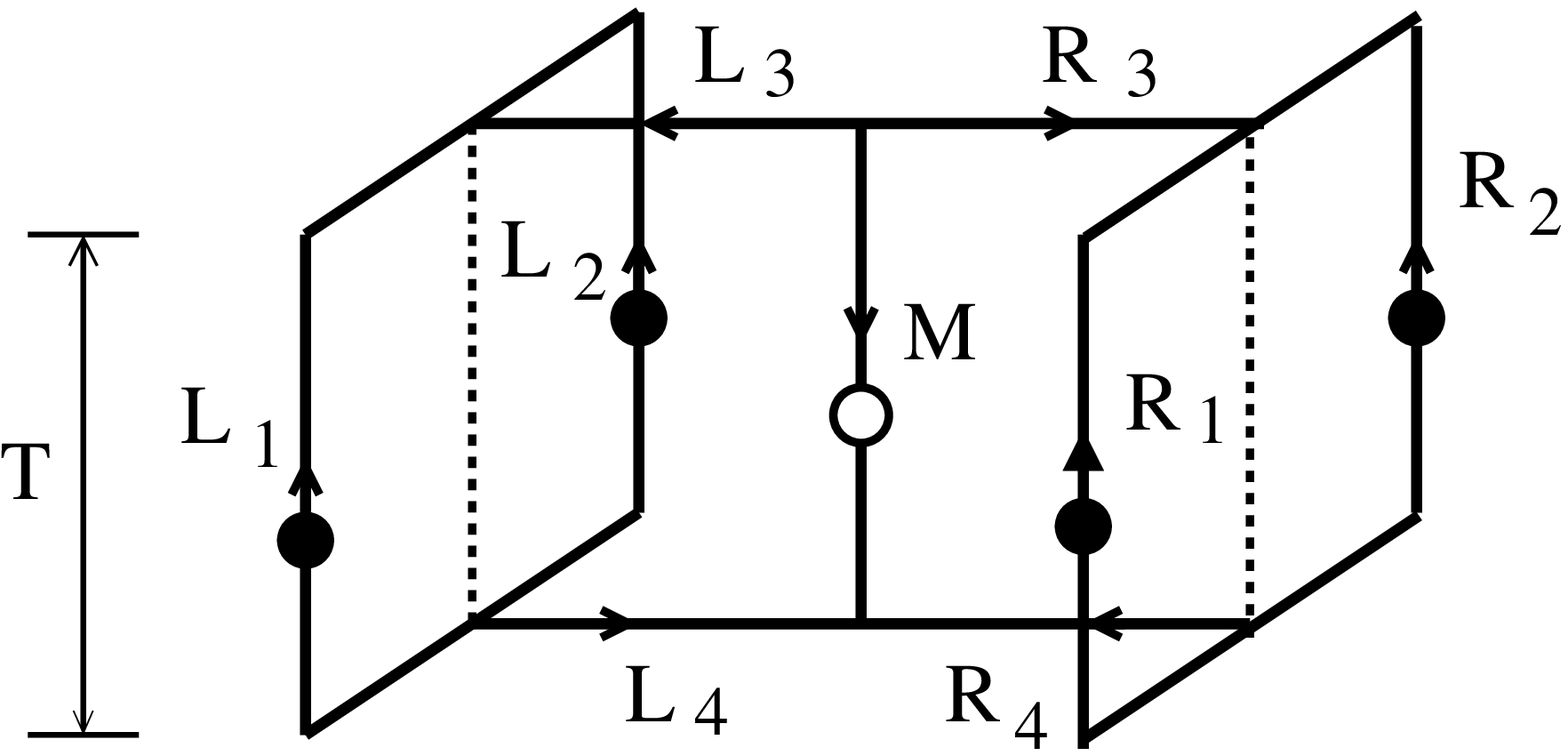} 
\includegraphics[height=3.3cm]{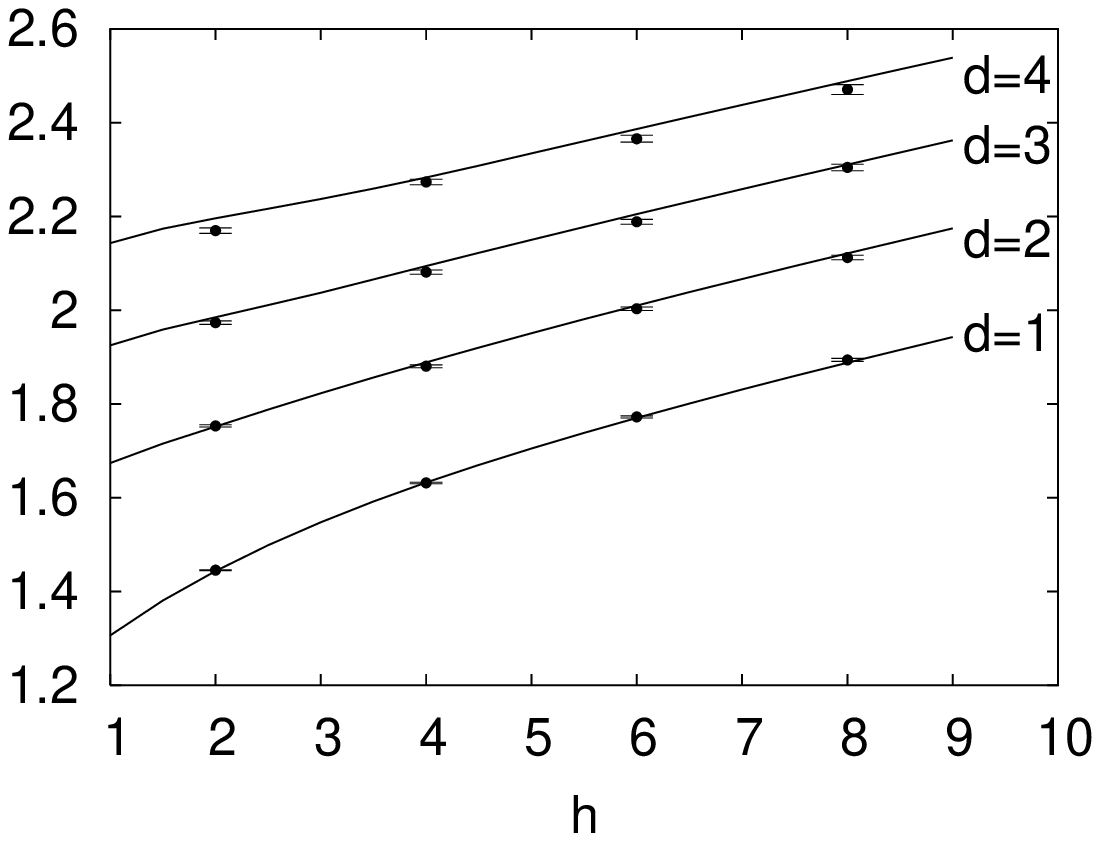}
\includegraphics[height=3.3cm]{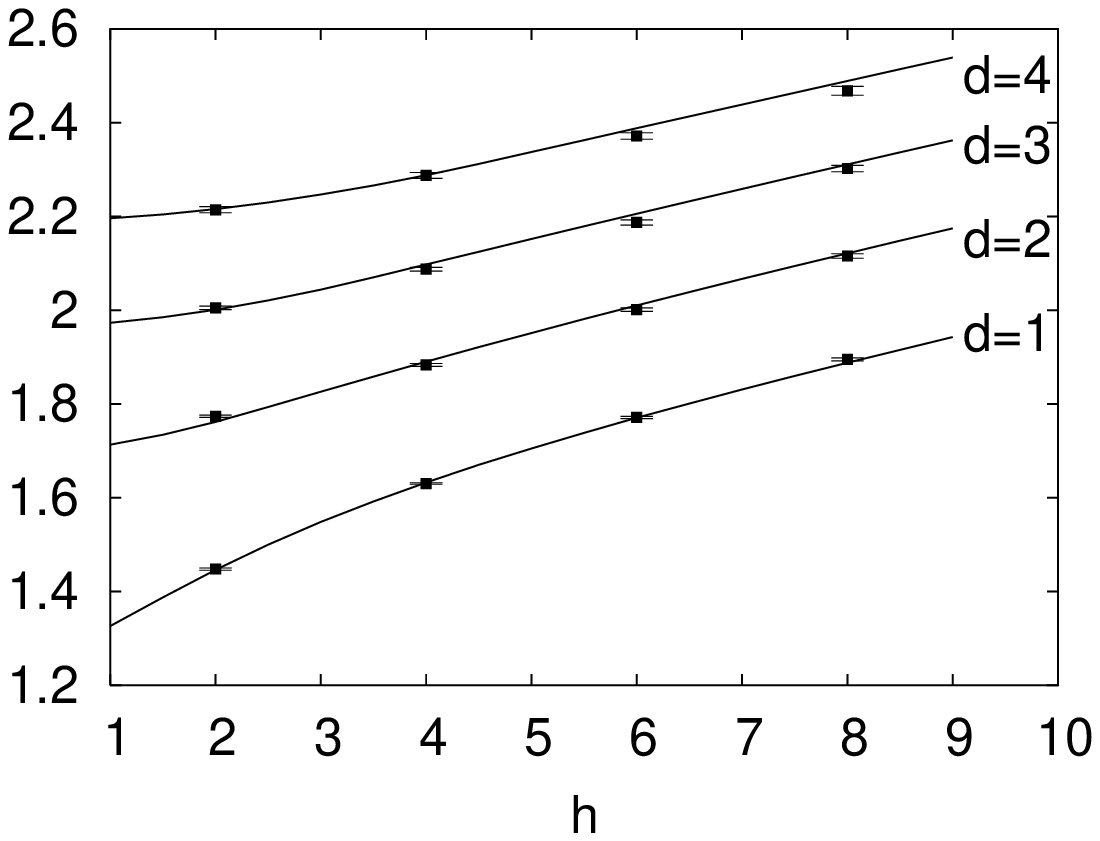} 
\vspace{-0.7cm}
\caption{
(a) The penta-quark (5Q) Wilson loop $W_{\rm 5Q}$ for the calculation of the 5Q potential $V_{\rm 5Q}$. 
(b) $V_{\rm 5Q}$ for planar configurations and (c) $V_{\rm 5Q}$ for twisted configurations.
The symbols denote the lattice QCD results, and the curves 
the OGE plus multi-Y Ansatz.}
\end{center}
\vspace{-1.5cm}
\end{figure}

\section{The Gluonic Excitation in the 3Q System}

Next, we study the gluonic excitation in lattice QCD \cite{STOI04,TS03,TS04}.
In the hadron physics, the gluonic excitation is one of the interesting phenomena 
beyond the quark model, and relates to the hybrid hadrons  
such as $q\bar qG$ and $qqqG$ in the valence picture \cite{P04}. 

For about 100 different patterns of 3Q systems, 
we perform the first study of the excited-state potential $V_{\rm 3Q}^{\rm e.s.}$
in SU(3) lattice QCD with $16^3\times 32$ at $\beta$=5.8 and 6.0 
by diagonalizing the QCD Hamiltonian in the presence of three quarks. 
The gluonic-excitation energy $\Delta E_{\rm 3Q} \equiv V_{\rm 3Q}^{\rm e.s.}-V_{\rm 3Q}^{\rm g.s.}$ 
is found to be about 1GeV in the hadronic scale \cite{STOI04,TS03,TS04}.
This result indicates that the lowest hybrid baryon $qqqG$ has a large mass of about 2 GeV.
\footnote{
Note that the gluonic-excitation energy of about 1GeV is rather large compared with 
the excitation energies of the quark origin. 
Therefore, for low-lying hadrons, the contribution of gluonic excitations 
is considered to be negligible, and the dominant contribution is brought 
by quark dynamics such as the spin-orbit interaction, 
which results in the quark model without gluonic modes \cite{STOI04,TS03,TS04}. 
}

\begin{figure}[h]
\vspace{-0.5cm}
\begin{center}
\includegraphics[height=3.15cm]{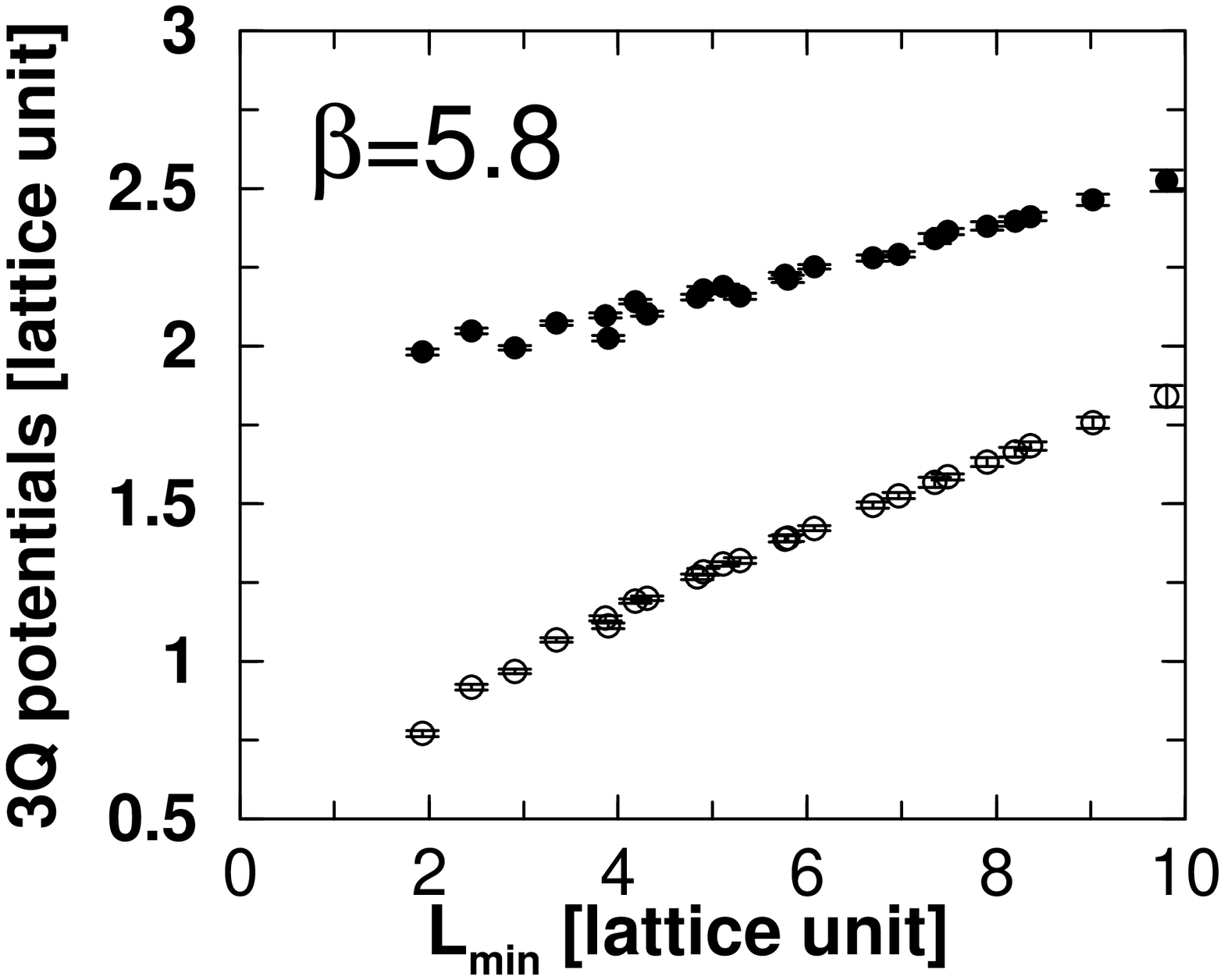}
\includegraphics[height=3.15cm]{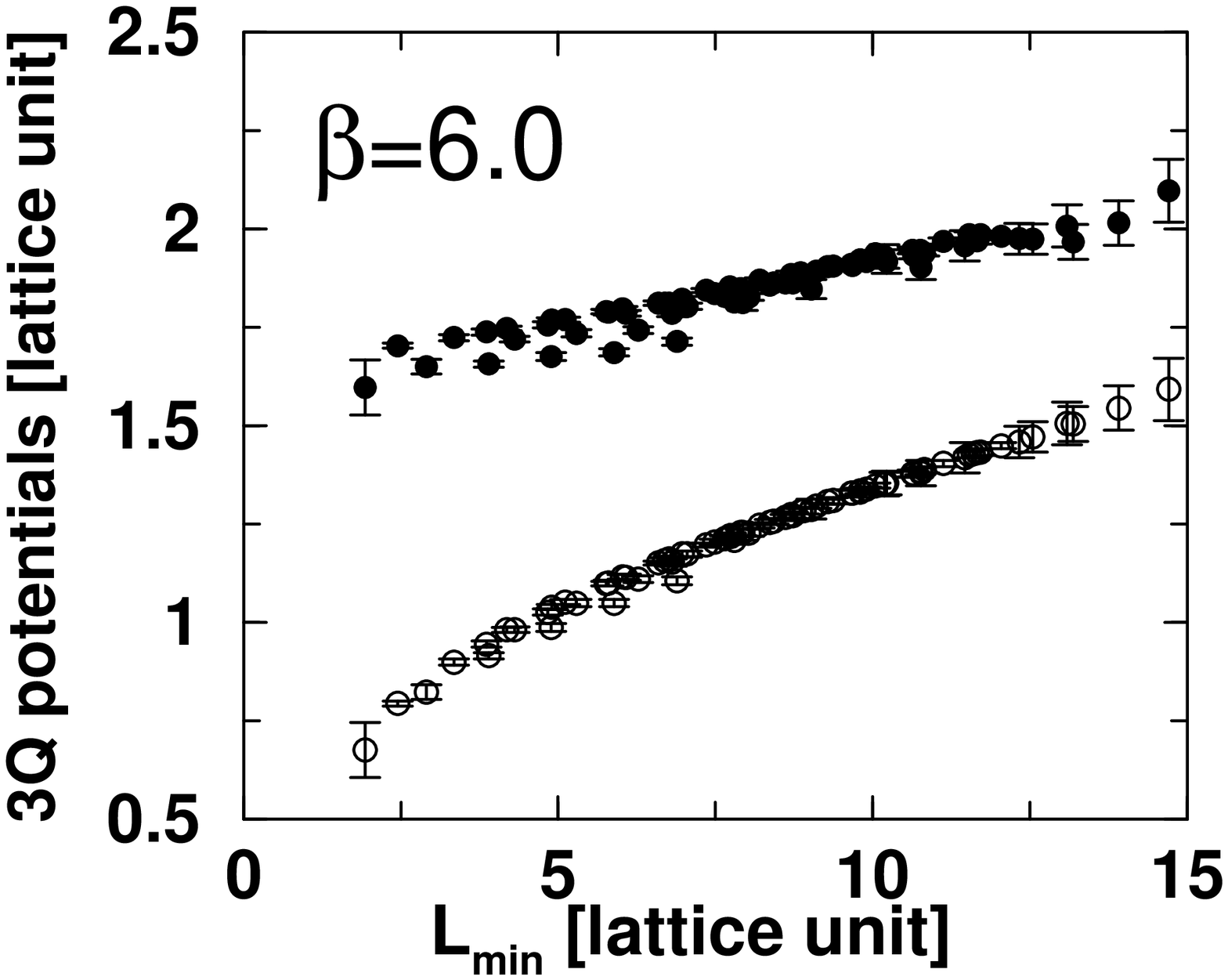}
\includegraphics[height=3.15cm]{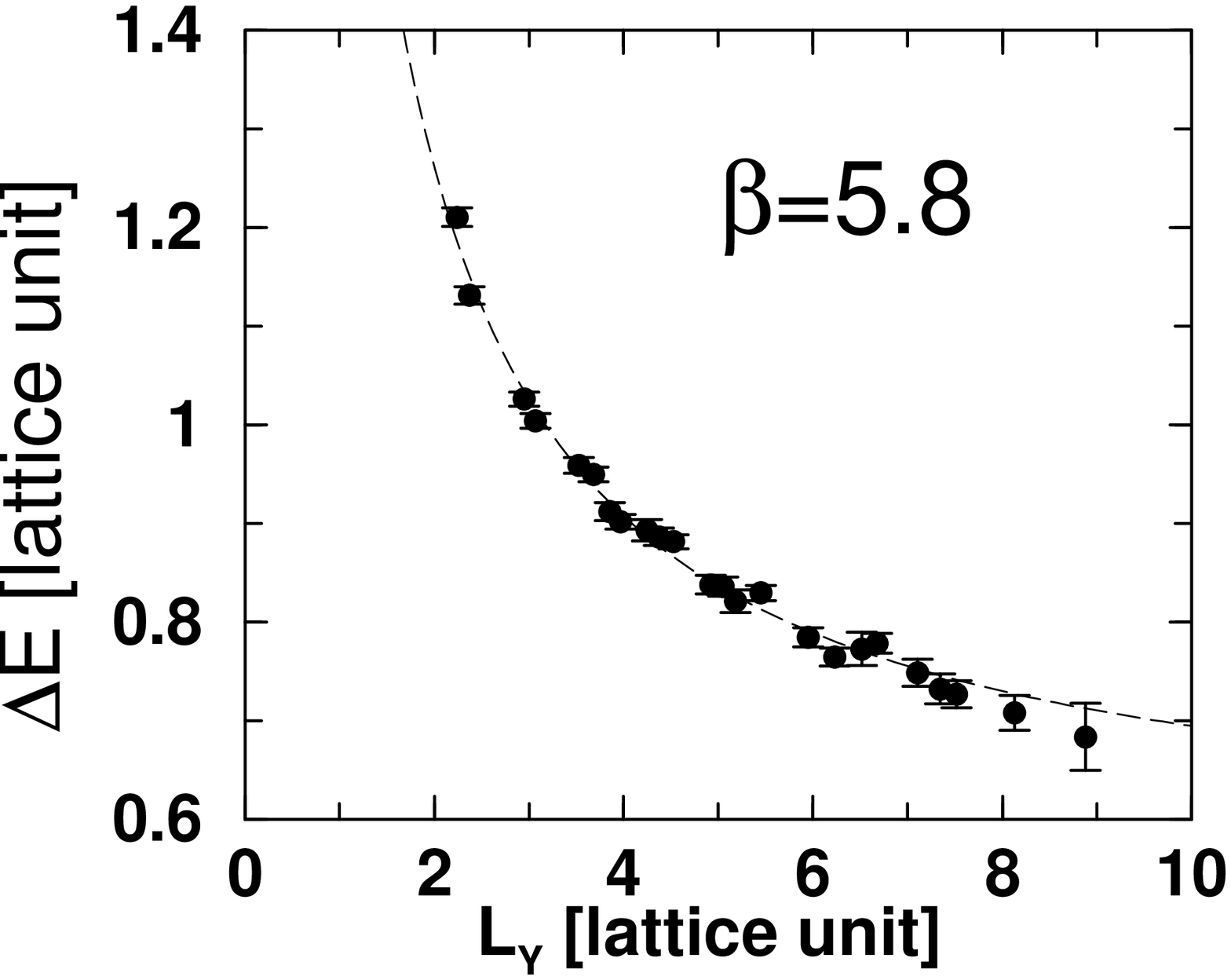}
\includegraphics[height=3.15cm]{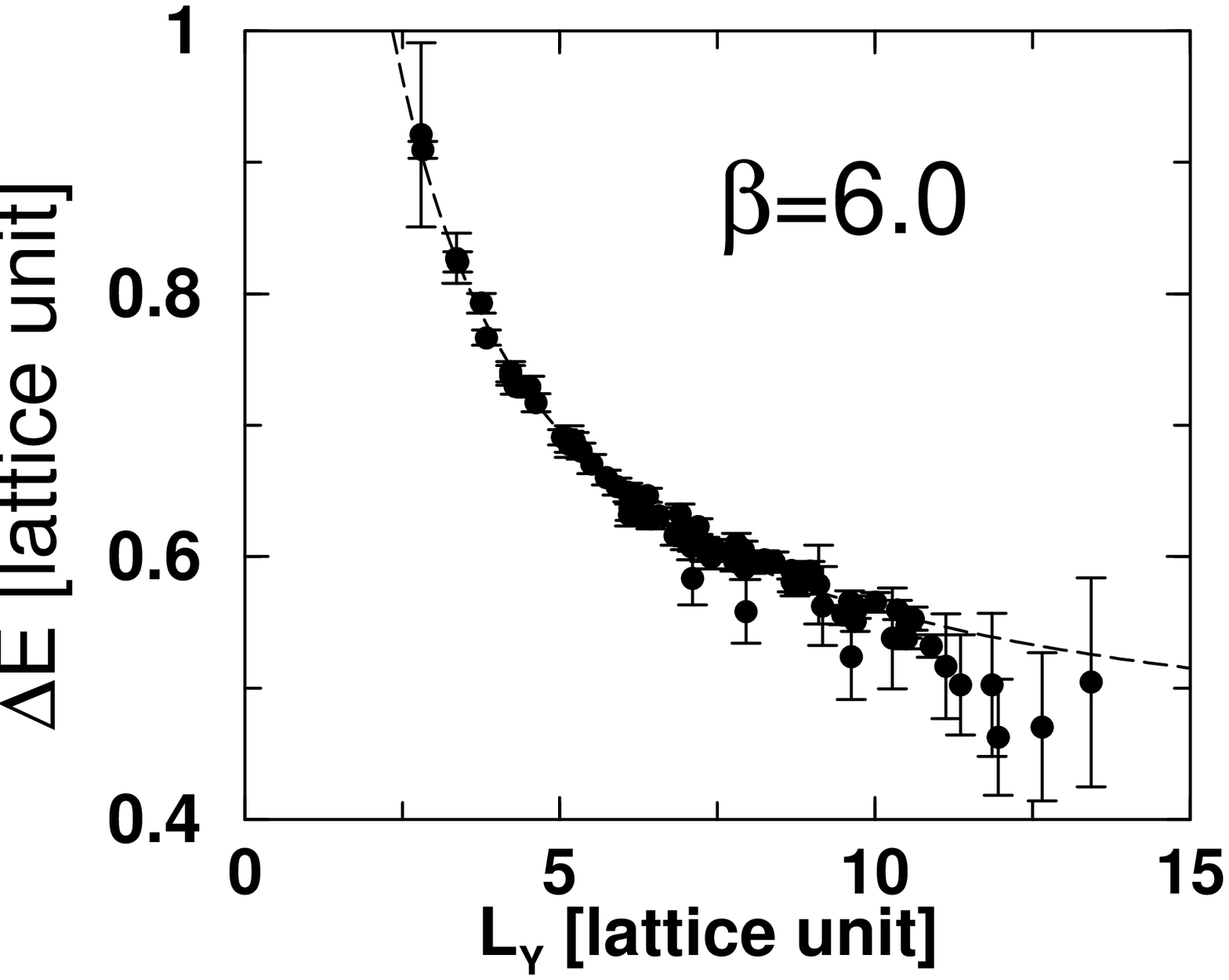}
\vspace{-0.75cm}
\caption{
(a) \& (b) The 1st excited-state 3Q potential 
$V_{\rm 3Q}^{\rm e.s.}$ and 
the ground-state 3Q potential $V_{\rm 3Q}^{\rm g.s.}$.
(c) \& (d) The gluonic excitation energy 
$\Delta E_{\rm 3Q} \equiv V_{\rm 3Q}^{\rm e.s.}-V_{\rm 3Q}^{\rm g.s.}$. 
The dashed curve denotes the ``inverse Mercedes Ansatz" \cite{STOI04,TS04}. 
}
\end{center}
\vspace{-1.3cm}
\end{figure}

\section{The QCD String Theory for the Penta-Quark Decay}

Our lattice QCD studies on the various inter-quark potentials 
indicate the flux-tube picture for hadrons, 
which is idealized as the QCD string model.
Here, we consider penta-quark dynamics, 
especially for its extremely narrow width, in the QCD string theory.

The ordinary string theory mainly describes open and closed strings corresponding to $\rm Q \bar Q$ mesons and glueballs, 
and has only two types of the reaction process:
the string breaking (or fusion) process and the string recombination process.

On the other hand, the QCD string theory describes also 
baryons and anti-baryons as the Y-shaped flux-tube, which  is 
different from the ordinary string theory.
Note that the appearance of the Y-type junction is peculiar to the QCD string theory with the SU(3) color structure.
Accordingly, the QCD string theory includes the third reaction process:
the junction (J) and anti-junction ($\rm \bar J$) par creation (or annihilation) process.
(Through this J-$\bar {\rm J}$ pair creation process, 
the baryon and anti-baryon pair creation can be described.)

\begin{figure}[h]
\vspace{-0.1cm}
\begin{center}
\includegraphics[width=7cm]{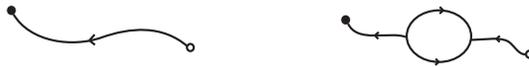}
\vspace{-0.5cm}
\caption{
The junction (J) and anti-junction ($\rm \bar J$) par creation (or annihilation) process.
}
\end{center}
\vspace{-1cm}
\end{figure}

As a remarkable fact in the QCD string theory, 
the decay/creation process of penta-quark baryons inevitably 
accompanies the J-$\bar {\rm J}$ creation \cite{BKST04,STOI04} 
as shown in Fig.5.
Here, the intermediate state is considered as a gluonic-excited state, since it clearly corresponds to 
a non-quark-origin excitation \cite{STOI04}. 

The lattice QCD study indicates that 
such a gluonic-excited state is a highly-excited state with the excitation energy above 1GeV.
Then, in the QCD string theory, 
the decay process of the penta-quark baryon near the threshold 
can be regarded as a quantum tunneling, 
and therefore the penta-quark decay is expected to be strongly suppressed.
This leads to a very small decay width of penta-quark baryons.

\begin{figure}[h]
\vspace{-0.2cm}
\begin{center}
\includegraphics[width=13cm]{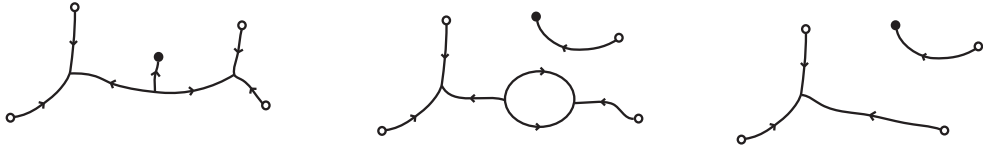}
\vspace{-0.8cm}
\caption
{A decay process of the penta-quark baryon in the QCD string theory.
The penta-quark decay process inevitably accompanies the J-$\bar {\rm J}$ creation, which is a kind of the gluonic excitation.
There is also a decay process via the gluonic-excited meson.
}
\end{center}
\vspace{-1.1cm}
\end{figure}

Now, we estimate the decay width of penta-quark baryons near the threshold 
in the QCD string theory. In the quantum tunneling as shown in Fig.5, 
the barrier height can be estimated as the gluonic-excitation energy $\Delta E \simeq$ 1GeV
of the intermediate state. 
The time scale $T$ for the tunneling process is expected to be the hadronic scale 
as $T =0.5 \sim 1{\rm fm}$, since $T$ cannot be smaller than the spatial size of the reaction area   
due to the causality. 
Then, the suppression factor for the penta-quark decay can be roughly estimated as 
$
|\exp(-\Delta E T)|^2 \simeq |e^{-1{\rm GeV} \times (0.5 \sim 1){\rm fm}}|^2 \simeq 
10^{-2}\sim 10^{-4}.
$
Note that this suppression factor $|\exp(-\Delta E T)|^2$
appears in the decay process of low-lying penta-quarks 
for both positive- and negative-parity states. 
For the decay of $\Theta^+(1540)$ into N and K, 
the decay width would be controlled by the Q-value, $Q \simeq 100{\rm MeV}$.
Considering the extra suppression factor of $|\exp(-\Delta E T)|^2$, 
we get a rough order estimate for the decay width of $\Theta^+(1540)$ as 
$
\Gamma[\Theta^+(1540)] \simeq Q \times |\exp(-\Delta E T)|^2 \simeq 1 \sim 10^{-2}{\rm MeV}.
$

\end{document}